\documentclass[twocolumn,showpacs,fleqn,nobibnotes]{revtex4}

\usepackage{amsmath}
\usepackage{graphicx}
\usepackage{float}
\usepackage{subfigure}

\def\lsim{\raise0.3ex\hbox{$<$\kern-0.75em\raise-1.1ex\hbox{$\sim$}}}
\def\gsim{\raise0.3ex\hbox{$>$\kern-0.75em\raise-1.1ex\hbox{$\sim$}}}

\newcommand{\rr}{\mbox{\boldmath $r$}}

\newcommand{\rb}{\mbox{\boldmath $b$}}
\newcommand{\rd}{\mbox{\boldmath $\Delta$}}

\begin{document}

\title{Photoproduction of $\rho^0$ mesons in ultraperipheral heavy ion collisions at energies available at the BNL Relativistic Heavy Ion Collider (RHIC) and CERN Large Hadron Collider (LHC)}
\author{V.P. Gon\c{c}alves
$^{a}$
and M.V.T. Machado $^{b}$ }

\affiliation{$^a$ Instituto de F\'{\i}sica e Matem\'atica, Universidade Federal de
Pelotas\\
Caixa Postal 354, CEP 96010-900, Pelotas, RS, Brazil.\\
$^b$ Centro de Ci\^encias Exatas e Tecnol\'ogicas, Universidade Federal do Pampa \\ Campus de Bag\'e, Rua Carlos Barbosa,  CEP 96400-700, Bag\'e, RS,  Brazil.}

\begin{abstract}
We investigate the photoproduction of $\rho$ mesons in ultraperipheral heavy ion collisions at RHIC and LHC energies in the dipole approach and within two phenomenological
models based on the the Color Glass Condensate (CGC) formalism.
We estimate the integrated cross section and rapidity distribution for  meson production and compare our predictions with the data from the STAR collaboration. In particular, we demonstrate that the total cross section at RHIC is strongly dependent on the energy behavior of the dipole-target cross section at low energies, which is not well determined in the dipole approach. In contrast, the predictions at midrapidities at RHIC and in the full rapidity at LHC are under theoretical control and can be used to test the QCD dynamics at high energies.

\end{abstract}

\maketitle

\section{Introduction}
The Large Hadron Collider (LHC) at CERN will start up in few months.  Currently, there is a great
expectation that LHC will discover the Higgs boson and whatever new physics beyond the Standard Model that may accompany it, such as
supersymmetry or extra dimensions \cite{lhc}. However, several questions remain open in the Standard Model, which will be probed in a new
kinematical regime at the LHC and will determine the background for new physics. In particular, the description of the high-energy regime of the Quantum
Chromodynamics (QCD) still is a subject of intense debate (For recent reviews see Ref. \cite{hdqcd}).


In recent years we have proposed the analysis of coherent collisions in hadronic interactions as an alternative way to study the QCD dynamics at high energies
\cite{vicmag_upcs,vicmag_hq,vicmag_mesons_per,vicmag_prd,vicmag_pA,vicmag_difper,vicmag_ane}. The basic idea in coherent  hadronic collisions is that
the total cross section for a given process can be factorized in
terms of the equivalent flux of photons into the hadron projectile and
the photon-photon or photon-target production cross section.
The main advantage of using colliding hadrons and nuclear beams for
studying photon induced interactions is the high equivalent photon
energies and luminosities that can be achieved at existing and
future accelerators (For a review see Ref. \cite{upcs}).
 Consequently, studies of $\gamma p$ interactions
at the LHC could provide valuable information on the QCD dynamics at
high energies.

The photon-hadron interactions in hadron-hadron collisions can be divided into exclusive and inclusive reactions. In the first case, a certain particle is produced while
the target remains in the ground state (or is only internally excited). On the other hand, in inclusive interactions the particle produced is accompanied by one or more
particles from the breakup of the target. The typical examples of these processes are the exclusive vector meson production and the inclusive heavy quark production,
described by the processes $\gamma h \rightarrow V h$ ($V = \rho, J/\Psi, \Upsilon$) and $\gamma h \rightarrow X Y$ ($X = c\overline{c}, b\overline{b}$), respectively. Recently, we have discussed both processes considering $pp$ \cite{vicmag_mesons_per,vicmag_prd,vicmag_ane}, $pA$ \cite{vicmag_pA} and $AA$ \cite{vicmag_hq,vicmag_mesons_per} collisions as an alternative to constrain the QCD dynamics at high energies (For recent reviews see Refs. \cite{vicmag_mpla,vicmag_jpg}). Our results demonstrate that their detection is feasible at the LHC. However, some of these results were obtained considering the  saturation model proposed by Golec-Biernat and W\"{u}sthoff (GBW) several years ago \cite{GBW}. That model has been improved recently considering the current state-of-the-art of  saturation physics: the Color Glass Condensate (CGC) formalism \cite{CGC,BAL,WEIGERT}. This fact motivates a revision  of some our previous estimates. In particular, in this paper  we revise  our predictions for $\rho$-meson photoproduction in coherent $AA$ collisions. We  show that the  rapidity distribution at RHIC is strongly dependent on the behavior of dipole-nucleus cross section at the threshold energy (and, therefore, model dependent), whereas the theoretical predictions based on CGC physics at LHC energy are under  control. Moreover, we demonstrate that the STAR Collaboration data for the rapidity distribution can be described by a phenomenological model which captures the basic properties of the quantum evolution in the CGC formalism.

This paper is organized as follows. In the next section we present a brief review of coherent hadron-hadron interactions, introducing the main formulas and  discussing the QCD dynamics at high energies. In particular, we present the  saturation models considered in the calculations. In Section \ref{resultados} we  compare our results  with the experimental data from STAR Collaboration at RHIC and present our predictions for the $\rho$ production at LHC. Finally, in Section \ref{conc} we summarize our main results and conclusions.

\section{Coherent $AA$ interactions and QCD dynamics at high energies}
\label{coerente}


Lets consider the hadron-hadron interaction at large impact parameter ($b > 2R_A$) at ultra relativistic energies. In this regime we expect the electromagnetic
interaction to be dominant. In  heavy ion colliders, the heavy nuclei give rise to strong electromagnetic fields due to the coherent action of all protons in the
nucleus, which can interact with each other. The photon emmited from the electromagnetic field
of one of the two colliding hadrons can interact with one photon of
the other hadron (two-photon process) or can interact directly with the other hadron (photon-hadron
process).  In what follows our main focus will be in photon - hadron processes.
Considering the requirement that  photoproduction
is not accompanied by hadronic interaction (ultra-peripheral
collision) an analytic approximation for the equivalent photon flux of a nuclei can be calculated, which is given by \cite{upcs}
\begin{eqnarray}
\frac{dN_{\gamma}\,(\omega)}{d\omega}= \frac{2\,Z^2\alpha_{em}}{\pi\,\omega}\, \left[\bar{\eta}\,K_0\,(\bar{\eta})\, K_1\,(\bar{\eta})+ \frac{\bar{\eta}^2}{2}\,{\cal{U}}(\bar{\eta}) \right]\,
\label{fluxint}
\end{eqnarray}
where
 $\omega$ is the photon energy,  $\gamma_L$ is the Lorentz boost  of a single beam and  $K_0(\bar{\eta})$ and  $K_1(\bar{\eta})$ are the
modified Bessel functions.
Moreover, $\bar{\eta}=2\omega\,R_A/\gamma_L$ and  ${\cal{U}}(\bar{\eta}) = K_1^2\,(\bar{\eta})-  K_0^2\,(\bar{\eta})$.
The Eq. (\ref{fluxint}) will be used in our calculations of $\rho$ meson production in $AA$ collisions at RHIC and LHC. 

The cross section for the $\rho$ meson photoproduction in a coherent  hadron-hadron collision is  given by,
\begin{eqnarray}
\sigma (AA\rightarrow AA\rho^0) = 2\,\int \limits_{\omega_{min}}^{\infty} d\omega \int dt \,\frac{dN_{\gamma}(\omega)}{d\omega}\,\frac{d\sigma}{dt} \left(W_{\gamma p},t\right)\,,
\label{sigAA}
\end{eqnarray}
where $\frac{d\sigma}{dt}$ is the differential cross section for the process    $(\gamma A \rightarrow \rho^0\,A)$, $\omega_{min}=M_{\rho}^2/4\gamma_L m_p$, $W_{\gamma p}^2=2\,\omega\sqrt{S_{\mathrm{NN}}}$  and
$\sqrt{S_{\mathrm{NN}}}$ is  the c.m.s energy of the
hadron-hadron system. The coherence condition limits the photon virtuality to very low values, which implies that for most purposes, they can be considered as real. In particular, for $PbPb$ collisions at LHC the Lorentz factor  is
$\gamma_L = 2930 $, giving the maximum c.m.s. $\gamma N$ energy
$W_{\gamma p} \approx 950$ GeV.  The factor two in Eq. (\ref{sigAA}) takes into account the fact that nuclei can act as both target and photon emitter. The experimental separation for such events is relatively easy, as photon emission is coherent over the entire nucleus and the photon is colorless we expect the events to be characterized by intact recoiled nuclei (tagged nuclei) and a two rapidity gaps pattern (For a detailed discussion see \cite{upcs}).


We describe the vector meson production in the color dipole frame, in which most of the energy is
carried by the hadron, while the  photon  has
just enough energy to dissociate into a quark-antiquark pair
before the scattering. In this representation the probing
projectile fluctuates into a
quark-antiquark pair (a dipole) with transverse separation
$\rr$ long after the interaction, which then
scatters off the hadron \cite{nik}.
In the dipole picture the  imaginary part of amplitude for vector meson  production reads  as (see Refs. \cite{nik,vicmag_mesons,KMW})
\begin{eqnarray}
 {\cal A}(x,\Delta) = 
\int dz d^2\rr  d^2\rb  e^{-i[\rb-(1-z)\rr].\rd}
 (\Psi_{V}^* \Psi_{\gamma}) \frac{d \sigma_{q\bar{q}}}{d^2\rb},
\label{sigmatot2}
\end{eqnarray}
where $\Psi^{\gamma}_{h, \bar{h}}(z,\,\rr)$ and $\Psi^{V}_{h,  \bar{h}}(z,\,\rr)$ are the light-cone wave function  of the photon  and of the  vector meson, respectively,
with the quark and antiquark helicities labeled by $h$ and $\bar{h}$. The variable $\rr$ defines the relative transverse
separation of the pair (dipole), $z$ $(1-z)$ is the longitudinal momentum fractions of the quark (antiquark), $\Delta$ denotes the transverse momentum lost by the
outgoing proton ($t = - \Delta^2$), and $x$ is the Bjorken variable. The transverse distance from the center of the target to one of the $q \bar{q}$ pair of the dipole is denoted by $\rb$. The differential dipole-target cross section is referred as $\frac{d \sigma_{q\bar{q}}}{d^2\rb}$. 
The photon wave functions appearing in Eq. (\ref{sigmatot2}) are well known in literature \cite{KMW}. For the meson wave function, we  consider the Gauss-LC  model \cite{KMW} (the results are insensitive to a different choice).  Moreover, we assume an effective light quark mass ($m_u = 0.14$ GeV). Finally, the differential cross section  for vector meson photoproduction is given by
\begin{eqnarray}
\frac{d\sigma}{dt} (\gamma A \rightarrow \rho^0 \,A) = \frac{1}{16\pi} |{\cal{A}}(x,\Delta)|^2\,(1 + \beta^2)\,,
\label{totalcs}
\end{eqnarray}
where $\beta$ is the ratio of real to imaginary parts of the scattering
amplitude. For the case of rho meson photoproduction, skewness corrections are not important and they are not taken  into account. A recent discussion on the vector meson production in $eA$ colliders and a detailed list of references about this subject can be found in Ref. \cite{mesons}.

\begin{figure}[t]
\includegraphics[scale=0.3]{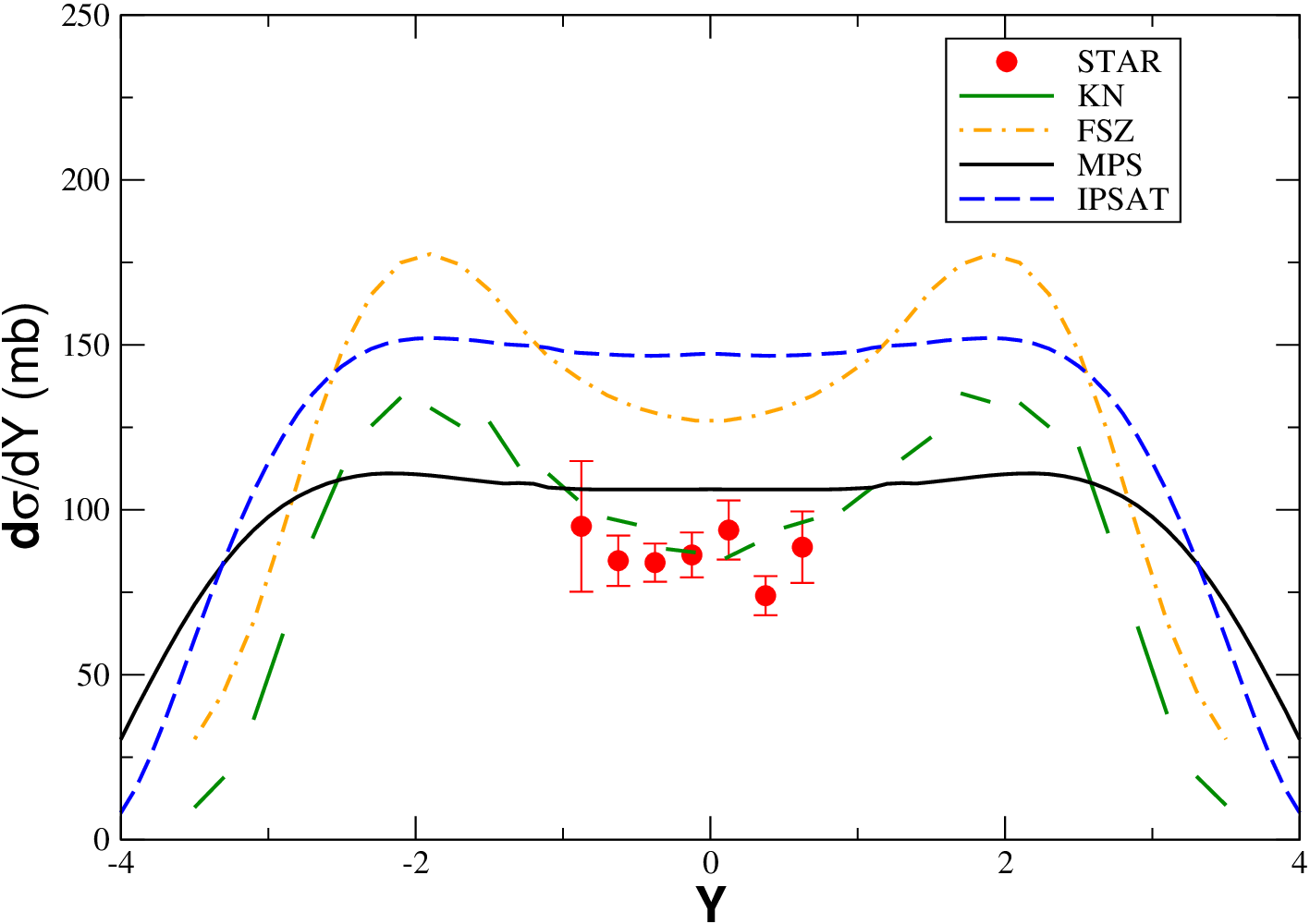}
 \caption{(Color online) Predictions for the rapidity distribution of $\rho^0$ photoproduction at RHIC energy considering distinct theoretical approaches. Data from STAR Collaboration \cite{STAR}.}
\label{fig:0}
\end{figure}

The differential dipole-target cross section contains all
information about the target and the strong interaction physics.
In the Color Glass Condensate (CGC)  formalism \cite{CGC,BAL,WEIGERT}, it  encodes all the
information about the
non-linear and quantum effects in the hadron wave function.
It can be obtained by solving an appropriate evolution
equation in the rapidity $y\equiv \ln (1/x)$ and its main properties are: (a) for the interaction of a small dipole ($r
\ll 1/Q_s$) the dipole-target satisfies the color transparency property ($\sigma_{q\bar{q}} \propto \rr^2$),  which characterizes that
the system is weakly interacting; (b) for a large dipole
($r \gg 1/Q_s$), the system is strongly absorbed.  This property is associate to the
large density of saturated gluons in the hadron wave function. In
our analysis we will  consider two saturation models generalized for the photon-nucleus interaction. The first one, generalizes the Iancu-Itakura-Munier (IIM) model \cite{IIM}, introducing the impact parameter dependence. In this model the differential dipole-nucleus cross section is given by \cite{armesto}
\begin{eqnarray}
\frac{d \sigma_{q\bar{q}}^{\mathrm{IIM}}}{d^2\rb} = 2 \left\{\, 1- \exp \left[-\frac{1}{2} A T_A(b)\, \sigma_{dip}^{\mathrm{IIM}}(x,r)  \right] \right\}\,, \nonumber
\end{eqnarray}
where $T_A(b)$ is the nuclear profile function (obtained from a 3-parameter Fermi distribution for the nuclear
density); the dipole-nucleon cross section is parametrized as \cite{IIM}
\begin{eqnarray}
\sigma_{dip}^{\mathrm{IIM}}(x,r) =\sigma_0 \,\left\{ \begin{array}{ll}
{\mathcal N}_0 \left(\frac{r^2Q_{s}^2}{4}\right)^{\gamma_{\mathrm{eff}}\,(x,\,r)}\,, & \mbox{for $rQ_s \le 2$}\,, \nonumber \\
 1 - \exp \left[ -c\,\ln^2\,(d\,rQ_s) \right]\,,  & \mbox{for $rQ_s  > 2$}\,,
\end{array} \right.
\label{CGCfit}
\end{eqnarray}
where the saturation scale is given by $Q_s(x) = (x_0/x)^{\lambda/2}$, with ${\mathcal N}_0 = 0.7$. The constants $c$ and $d$ are determined from the condition that $\sigma_{dip}$ and its derivative with respect to $rQ_s$ are continuous at $rQ_s = 2$. The expression for $rQ_s > 2$  (saturation region)   has the correct functional form, as obtained  from the theory of the Color Glass Condensate (CGC) \cite{CGC}. Moreover,  $\gamma_{\mathrm{eff}}\, (x,\,r)= \gamma_{\mathrm{sat}} + \frac{\ln (2/rQ_s)}{\kappa \,\lambda \,y}$ is the   effective anomalous dimension, which determines the behavior of the dipole cross section in the color transparency regime ($rQ_s < 2$) and introduces evolution effects not included in the GBW model \cite{GBW}. The parameters of model are determined from a fit to $F_2$ data. The IIM model can be considered a phenomenological model for the quantum limit of the CGC formalism, since it  captures the basic properties of the quantum evolution, describing both the bremsstrahlung limit of linear small-$x$ evolution (BFKL equation) as well as nonlinear renormalization group at high parton densities (very small-$x$).
For comparison we also consider the IP-SAT model generalized for nuclei target proposed in Ref. \cite{Lappi}, which is a   a phenomenological model for the classical limit of the CGC. In this model, the differential dipole-nucleus cross section takes the form
\begin{eqnarray}
\frac{d \sigma_{q\bar{q}}^{\mathrm{IP-SAT}}}{d^2\rb} = 2\, \left[1 - \exp\left( - \frac{\pi^2r²}{2\,N_c} \alpha_s(\mu^2)\,xg(x,\mu^2) AT(\rb)\right)\right], \nonumber
\label{nkmw}
\end{eqnarray}
where the scale $\mu^2$ is related to the dipole size $\rr$ by $\mu^2 = 4/r^2 + \mu_0^2$ and the gluon density is evolved from a scale $\mu_0^2$ up to $\mu^2$ using LO DGLAP evolution without quarks, assuming that the initial gluon density is given by $xg(x,\mu_0^2) = A_g \, x^{-\lambda_g}\,(1 - x)^{5.6}$. The values of the parameters $\mu_0^2$, $A_g$ and $\lambda_g$ are determined from a fit to $F_2$ data.  The parameter set used in our calculations is the one presented in the first line of Table III of \cite{KMW}. It is important to emphasize that both models describe quite well the small-$x$ HERA data  ($x\le 0.01$). Therefore, the study of  other observables which are strongly dependent on $\frac{d \sigma_{q\bar{q}}}{d^2\rb}$ is very important to constrain the underlying QCD dynamics at high energies.

\begin{figure*}[t]
\includegraphics[scale=0.3]{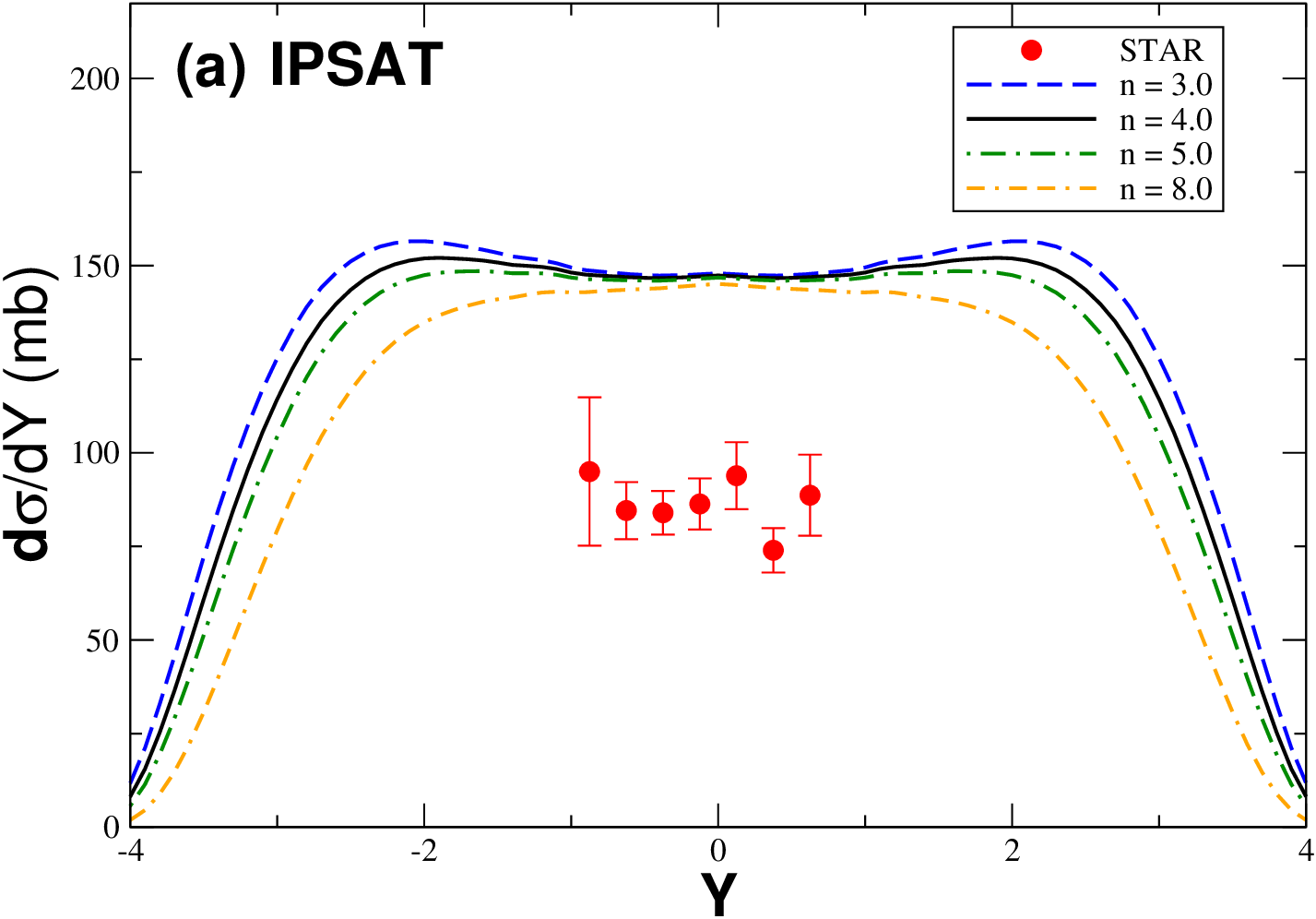} 
\hspace{0.4cm}
\includegraphics[scale=0.3]{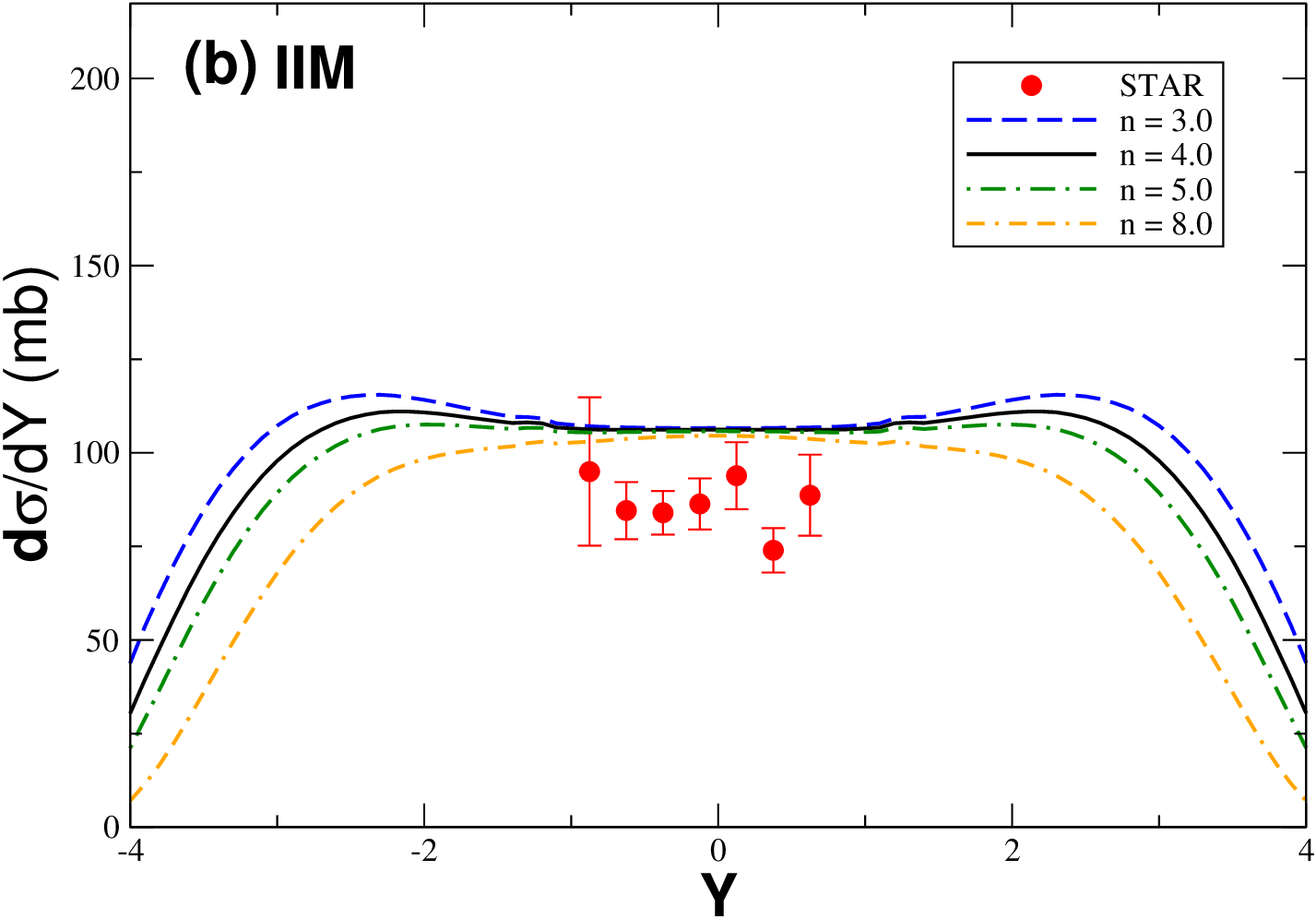}
 \caption{(Color online) The rapidity distribution of $\rho^0$ meson photoproduction at RHIC energy ($\sqrt{s}=200$ GeV) for IP-SAT  (left panel) and IIM  (right panel) models. Data from STAR Collaboration \cite{STAR}.}
\label{fig:1}
\vspace{1cm}
\end{figure*}

An important comment is in order here. The description of the vector meson production using the color dipole picture is justified at small values of $x$ ($x \le 0.01$) and large $x$ contributions are not included.  As we will show in next section, in the kinematical region of large rapidities at RHIC,  the dipole-cross section is probed in the large-$x$ region. In order to take this fact into account, we will multiply the differential dipole-nucleus cross sections presented above by a threshold factor, $(1-x)^n$, as performed in previous studies \cite{vicmag_ane,vicmag_mesons}. Notice that in the IP-SAT model, a large-$x$ correction is already included in the proton gluon distribution function. We discuss the consequence of this phenomenological procedure in what follows.

\section{Results}
\label{resultados}

Lets calculate the rapidity distribution and total cross section for the $\rho^0$ meson photoproduction in  coherent $AA$ collisions.
The distribution on rapidity $y$ of the produced final state can be directly computed from Eq. (\ref{sigAA}), by using its  relation with the photon energy $\omega$, i.e. $y\propto \ln \, (2 \omega/M_{\rho})$.  Explicitly, the rapidity distribution is written down as,
\begin{eqnarray}
\frac{d\sigma \,\left[A + A \rightarrow   A \otimes \rho^0 \otimes A \right]}{dy} = \omega \frac{dN_{\gamma} (\omega )}{d\omega }\,\sigma_{\gamma A \rightarrow \rho^0\, A}\left(\omega \right)\,
\label{dsigdy}
\end{eqnarray}
where $\otimes$ represents the presence of a rapidity gap. Consequently, given the photon flux, the rapidity distribution is thus a direct measure of the photoproduction cross section for a given energy. 


\begin{figure}
\includegraphics[scale=0.3]{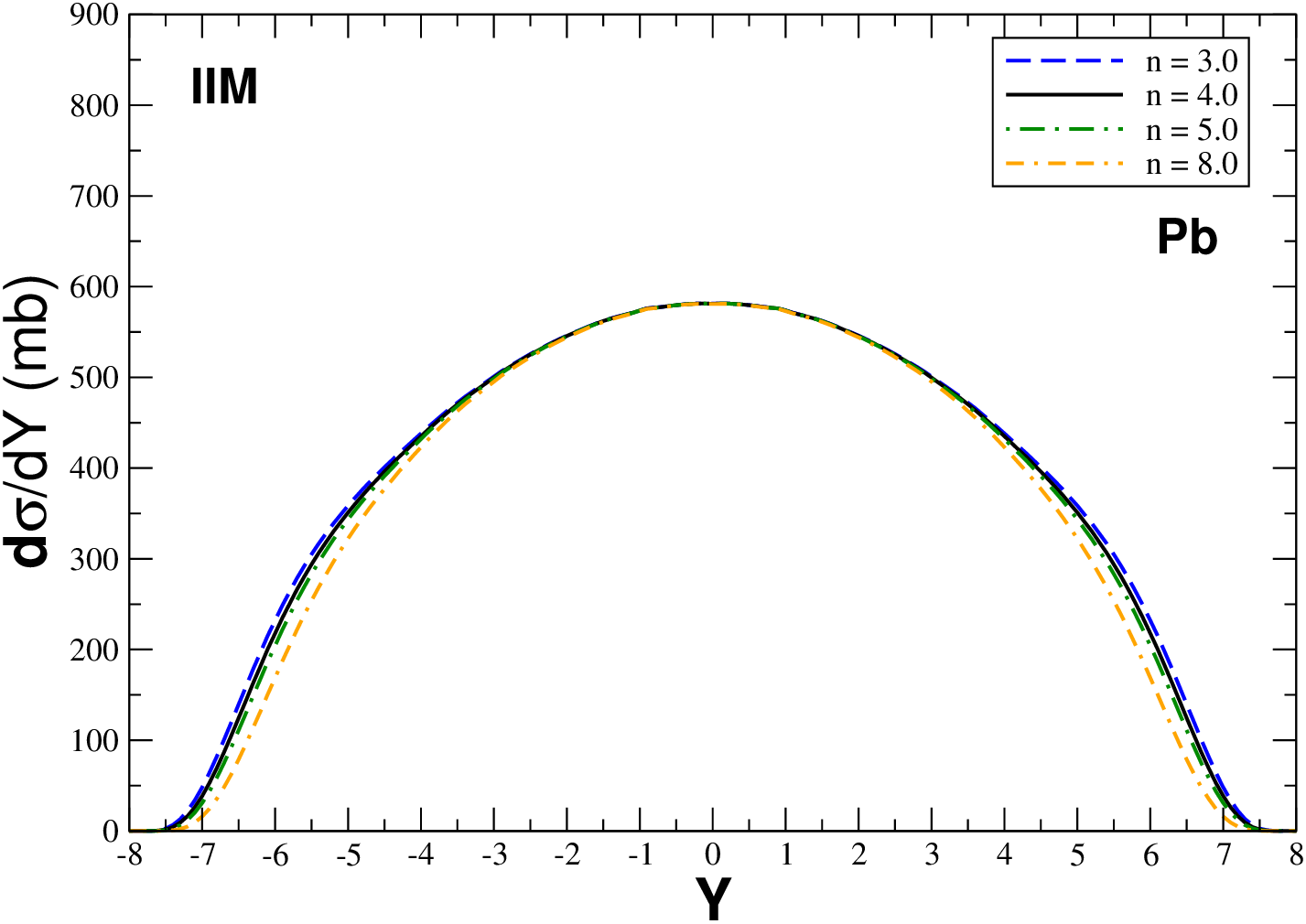} 
 \caption{(Color online) Dependence on the threshold factor of the  rapidity distribution of $\rho^0$ photoproduction at the LHC (PbPb collisions and $\sqrt{s}=5.5$ TeV) considering the IIM model.}
\label{fig:2}
\end{figure}

Initially, lets investigate  the $\rho^0$ photoproduction at RHIC energies.  The present calculation update our previous calculation \cite{vicmag_mesons_per} and allow us to compare our predictions with those from Refs.  \cite{klein_prc,strikman} and the experimental data  presented in Ref. \cite{STAR}. In Fig. \ref{fig:0} we compare our predictions with those from Klein and Nystrand (KN) \cite{klein_prc} and Frankfurt, Strikman and Zhalov (FSZ) \cite{strikman}, which are based on distinct theoretical approaches (For a detailed discussion of these models see \cite{vicmag_mesons_per}). We verify that the IP-SAT model overestimates the STAR data at mid rapidity, whereas the IIM model seems to be  consistent with the overall normalization of the experimental measurements. In comparison with the KN and FSZ predictions, the predictions of the saturation models are not strongly suppressed at mid rapidities and $\sqrt{s} = 200 $ GeV. Such  suppression occurs only at smaller energies. 

The results presented in Fig. \ref{fig:0} were obtained using $n = 4$ in the threshold factor.
In order to verify the dependence of our predictions in the model used for the low energy regime, which is not under theoretical control in the dipole approach, we show in detail in Fig. \ref{fig:1} the dependence on the threshold factor, $(1-x)^n$, taking different values for the index $n$ ($n=3, 4, 5, 8$). The form of the threshold factor was  inspired by the quark counting rules. The effect is completely negligible at central rapidity, whereas it is important at forward rapidities. The dependence on $n$ starts to be very important for rapidities $|y|>2$. It implies that our predictions  in the full rapidity range are strongly  dependent on the model used to describe the low energy regime. 
This fact has important consequences when computing the total exclusive cross section at RHIC energy. In contrast, we have that at LHC energies the threshold factor is negligible for most of rapidity range, being of marginal importance for very forward rapidities, as can be observed in Fig. \ref{fig:2}, where  we present the IIM  predictions for  the rapidity distribution for $\rho^0$ photoproduction in PbPb collisions at LHC ($\sqrt{s}=5500$ GeV). A similar behavior  is observed using the IP-SAT model.  Therefore, the color dipole model is a suitable approach to describe $\rho$ meson production at LHC. 

\begin{figure}
\includegraphics[scale=0.3]{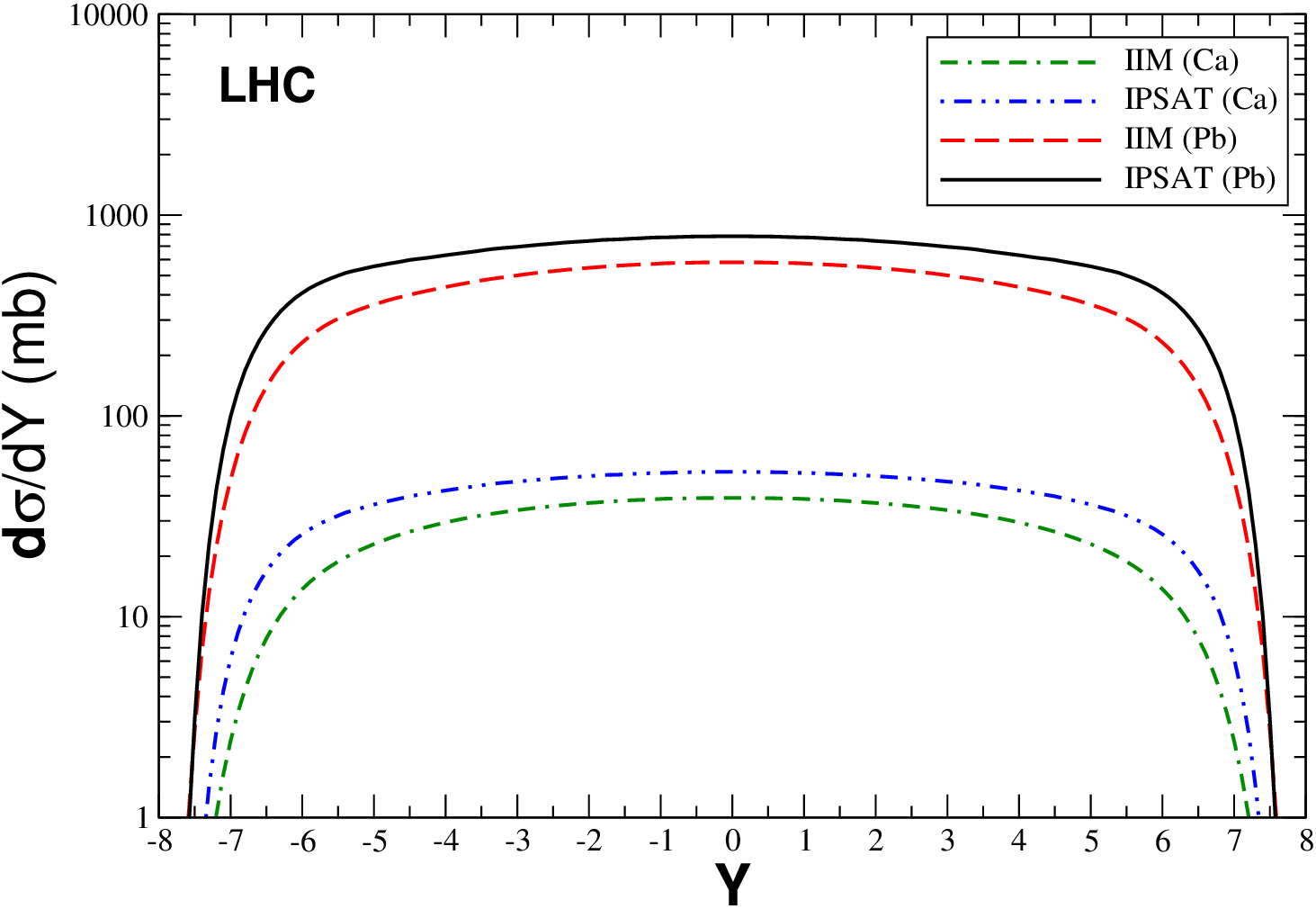}
 \caption{(Color online) Predictions for the rapidity distribution of $\rho^0$ photoproduction in PbPb and CaCa collisions at the LHC. The results for IP-SAT and IIM models are presented.}
\label{fig:3}
\end{figure}
\begin{figure*}[t]
\begin{tabular}{c c}
\includegraphics[scale=0.3]{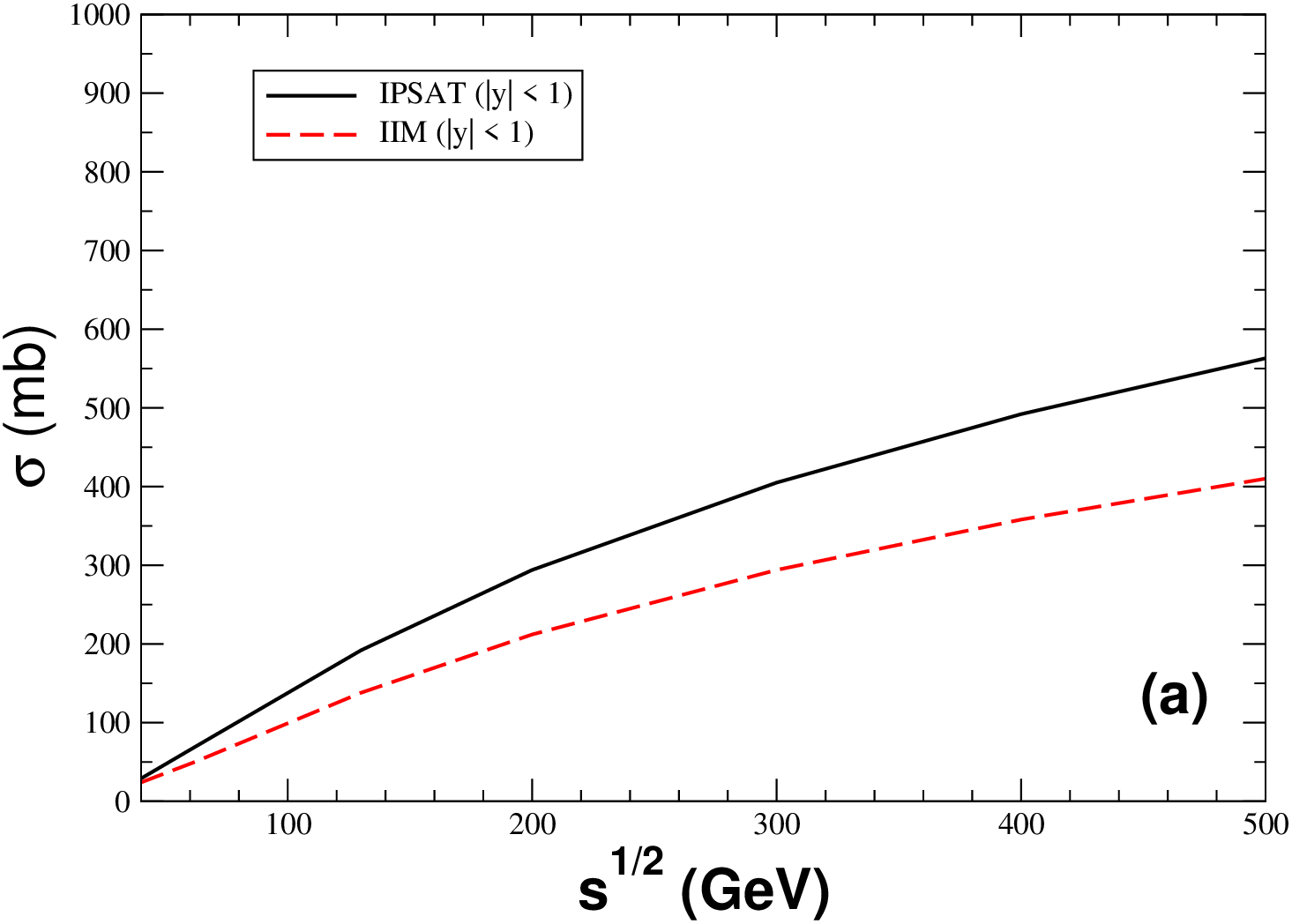} &
\includegraphics[scale=0.3]{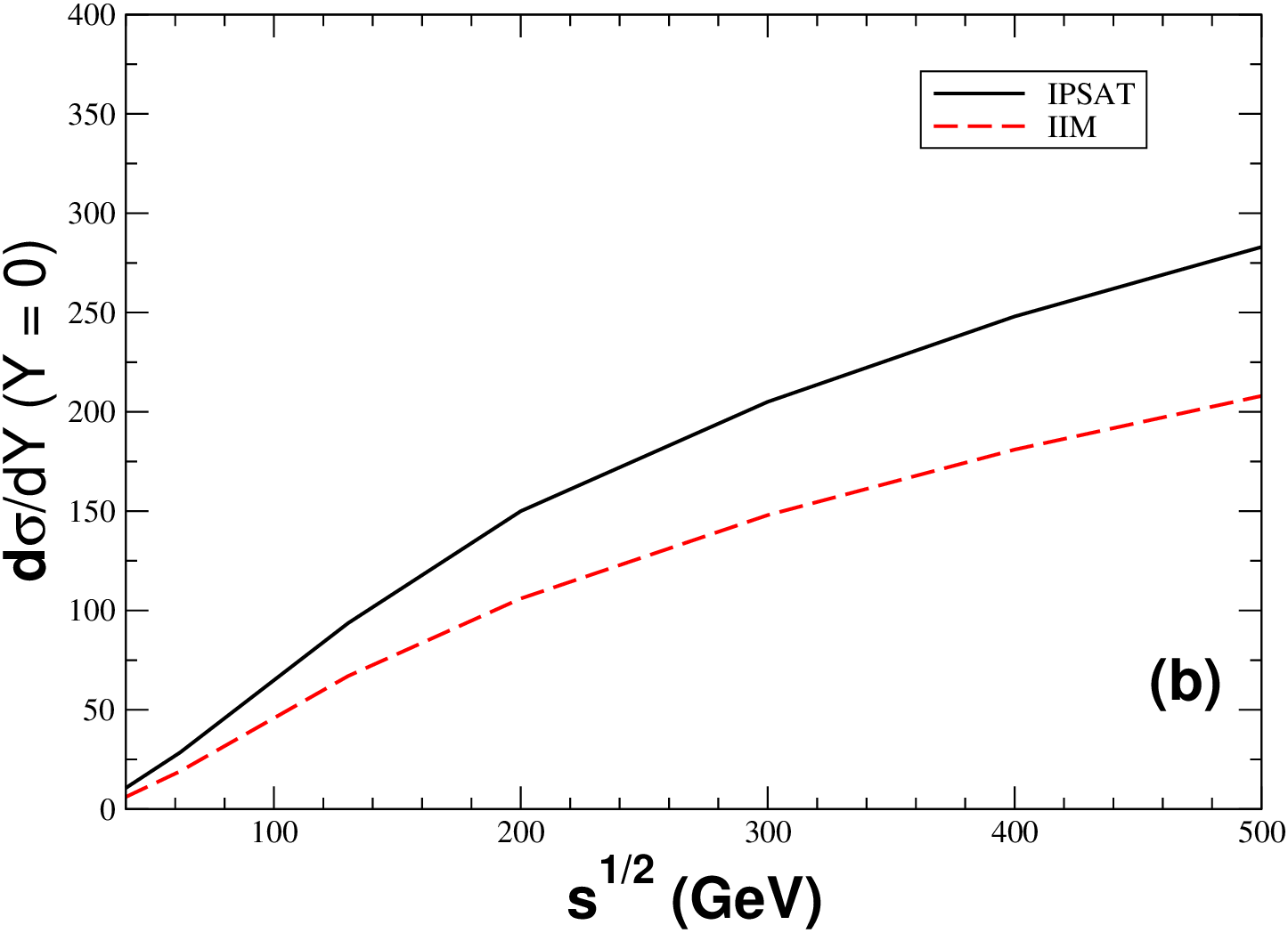}
\end{tabular}
 \caption{(Color online) The total exclusive cross section as a function of energy in the range $50\leq\sqrt{s}\leq 500$ GeV (left panel) for the IP-SAT and IIM models (using cut $|y|<1$). The differential cross section $d\sigma/dy$ at central rapidity $y=0$ as a function of energy (right panel).}
\label{fig:4}
\end{figure*}

In Fig. \ref{fig:3} we present our predictions for the rapidity distribution for the LHC energy considering PbPb and CaCa collisions. The results for both models are shown for the full rapidity range. The rapidity distribution at LHC probes a large interval of photon-nucleus center of mass energy since $W^2_{\gamma A}\simeq M_{\rho}\,\sqrt{s}e^{\pm y}$, which corresponds to very small $x\simeq M_{\rho}e^{-y}/\sqrt{s}$. Therefore, its  experimental analyzes can be useful to determine the QCD dynamics. As already observed  at RHIC energies,  the IIM predictions are smaller than the IP-SAT one for collisions with both nuclei. However, the difference between the predictions of the models increases with the energy, as observed in the left panel of Fig. \ref{fig:4} where we present the energy dependence of the total cross section calculated  assuming the cut $|y|<1$. The reason to impose this kinematical cut is based on the  sensitivity to the large-$x$ correction to the dipole cross section. The plot is motivated by recent studies presenting the energy dependence of the $\rho^0$ photoproduction cross section for the RHIC energies.  The large difference between the predictions is directly associated to the distinct energy dependence predicted by the IIM and IP-SAT model for the photo-nucleus cross section. In Fig. \ref{fig:4} (right panel) we present our predictions for the energy dependence of $d\sigma/dy$ at central rapidity ($y=0$), which is not dependent on the model used to describe the large $y$ region. A similar behavior is also observed for this quantity, which implies that its experimental study can be useful to discriminate between the saturation models. Our predictions for the rapidity distribution at $y = 0$ and LHC energies (See Fig. \ref{fig:3}) can be compared with those presented in Refs. \cite{klein_prc,strikman}. In comparison with Ref. \cite{klein_prc}, our prediction for $\rho^0$ photoproduction in CaCa collisions is a factor $\ge 3$ larger, which is directly associated to the distinct   energy dependence among the models. In Ref. \cite{klein_prc} it is assumed that the $\rho^0$ photoproduction is described by soft physics, which implies a smooth energy dependence. In contrast, in the models based on CGC physics, it is expected that at high energies the saturation scale determines the QCD dynamics implying a stronger energy dependence.  In comparison with Ref. \cite{strikman}, our predictions for $d\sigma/dy (y=0)$ for  $\rho^0$ photoproduction in PbPb collisions are similar.

\begin{table}
\begin{center}
\begin{tabular} {||c|c|c||}
\hline
\hline
$A$& {\bf IIM} &  {\bf IP-SAT}  \\
\hline
\hline
Ca &  395 mb (16985.0) & 572 mb  (24600.0) \\
\hline
 Pb &  5977 mb (2510.0) & 8596 mb  (3610.0) \\
\hline
\hline
\end{tabular}
\end{center}
\caption{\ The integrated cross section (events rate/second) for the $\rho^0$ photoproduction  in $AA$  collisions at LHC energies and different nuclei ($A$ = Pb and Ca).}
\label{tabrho}
\end{table}

We are now ready to compute the total exclusive cross section, Eq. (\ref{sigAA}), and production rates for LHC energies  considering the distinct phenomenological models. The results are presented in Table \ref{tabrho}.  At LHC we assume the  design luminosity ${\cal L}_{\mathrm{LHC}} = 0.42 \, (43.0)$ mb$^{-1}$s$^{-1}$ for PbPb (CaCa) collisions.
The corresponding cross sections are very large, which implies that the experimental analyzes of this process at LHC should be feasible. In particular, as the $\rho^0$ meson photoproduction is an exclusive reaction, the separation of the signal from hadronic background would be relatively clear. Namely, the characteristic features  are low $p_T$ $\rho^0$ spectra and a double rapidity gap pattern. Moreover, the detection (Roman pots) of the scattered nuclei can be an additional useful feature.  Moreover, in comparison with the hadroproduction of vector mesons, the event multiplicity for photoproduction interactions is lower, which implies that it may be used as a separation factor between these processes (For a detailed discussion see Refs. \cite{vicmag_mesons_per,klein_prc}). In comparison with previous results \cite{vicmag_mesons_per,klein_prc,strikman} our predictions are smaller than those presented in \cite{strikman}, obtained using the Generalized Vector  Dominance Model (GVDM), with the IIM one being a factor $\approx 1.6$ smaller. In contrast, they are a factor $\ge 3$  larger than those presented in \cite{klein_prc}, where a VDM plus a classical mechanical calculation for nuclear scattering was used in order to calculate the rapidity distribution and cross sections (For a detailed comparison of these models with the dipole approach see Section IV in Ref. \cite{vicmag_mesons_per}).  In comparison with our previous results \cite{vicmag_mesons_per} the IIM and IP-SAT predictions for $PbPb$ collisions are smaller than those obtained using the GBW model, which is directly associated to the distinct energy dependence of the dipole-target cross sections.

\section{Conclusions} 
\label{conc}
 
The QCD dynamics at high energies is of utmost importance for building a realistic description of $AA$ collisions at LHC. In this limit QCD evolution leads to a system with high gluon density, denoted Color Glass Condensate (CGC). If such system there exists at high energies it can be proven in coherent $AA$ collisions at the LHC. In this paper we have analyzed the $\rho^0$ meson photoproduction considering the dipole approach and most recent phenomenological models based in CGC physics.  We shown that the predictions of the dipole approach for rapidity distribution at RHIC are strongly dependent on the behavior of dipole-nucleus cross section at threshold energy (and, therefore, model dependent). It implies that this approach is not able to give reliable predictions for the total cross section at RHIC. However, at central rapidities ($y<|1|$), where are the experimental data of the STAR Collaboration, our prediction are not strongly dependent on the model used at low energies. At this kinematical region we demonstrate that the STAR data can be described by the IIM model, which captures the basic properties of the quantum evolution in the CGC formalism. Furthermore, we demonstrate that  the predictions for the rapidity distribution and total cross section based on CGC physics are under theoretical control  at the LHC energy.  Moreover, we predict very large cross sections and event rates, which implies the experimental study of this process is feasible at LHC. Consequently, we believe that it can be used to study the QCD at high energies.

\begin{acknowledgments}
VPG would like to thanks E.G.S. Luna for useful discussions. This work was  partially financed by the Brazilian funding agencies CNPq and FAPERGS.
\end{acknowledgments}

\end{document}